\documentclass[twocolumn,showpacs,superscriptaddress,preprintnumbers,prd]{revtex4}
\usepackage{graphicx,amsmath,dcolumn}
\begin{document}
\preprint{SAGA-HE-205-04}
\title{Nuclear modification difference between $u_v$ and $d_v$ distributions \\
       and its relation to NuTeV $\sin^2 \theta_W$ anomaly}
\author{M. Hirai}
\email{mhirai@post.kek.jp}
\affiliation{Institute of Particle and Nuclear Studies, KEK,
              1-1, Ooho, Tsukuba, Ibaraki, 305-0801, Japan}
\author{S. Kumano\footnote{Address after Jan. 1, 2005: 
                  Institute of Particle and Nuclear Studies, KEK,
                   1-1, Ooho, Tsukuba, Ibaraki, 305-0801, Japan.
                   URL: http://research.kek.jp/people/kumanos/}}
\email{shunzo.kumano@kek.jp}
\affiliation{Department of Physics, Saga University,
         Saga, 840-8502, Japan}
\author{T.-H. Nagai}
\email{tnagai@post.kek.jp}
\affiliation{Department of Physics, Saga University,
             Saga, 840-8502, Japan}
\date{December 18, 2004}
\begin{abstract}
\vspace{0.2cm}
We investigate a possible nuclear correction to the NuTeV measurement of
the weak-mixing angle $\sin^2 \theta_W$. In particular, a nuclear modification
difference between $u_v$ and $d_v$ distributions contributes to the NuTeV
measurement with the iron target. First, the modification difference
is determined by a $\chi^2$ analysis so as to reproduce nuclear data
on the structure function $F_2$ and Drell-Yan processes. Then, taking
the NuTeV kinematics into account, we calculate a contribution
to the $\sin^2 \theta_W$ determination. In addition, its uncertainty is
estimated by the Hessian method. Although the uncertainty becomes comparable
to the  NuTeV deviation, the effect is not large enough to explain the whole
NuTeV $\sin^2 \theta_W$ anomaly at this stage. However, it is difficult
to determine such a nuclear modification difference, so that we need
further investigations on the difference and its effect on the NuTeV anomaly.
\end{abstract}
\maketitle

\section{Introduction}\label{intro}
\setcounter{equation}{0}
\vspace{-0.20cm}

Weak-mixing angle $\sin^2 \theta_W$ is one of fundamental constants,
and there are various experimental methods for
determining it \cite{pdg04}. They include experiments on
atomic parity violation, left-right and forward-backward asymmetries
in electron-positron annihilation, $W$ and $Z$ masses, elastic and
inelastic neutrino scattering, parity-violating electron scattering,
and polarized M$\o$ller scattering.

The NuTeV collaboration announced that their neutrino-nucleus scattering
data indicate a significant deviation from the standard model.
If the neutrino-nucleus scattering data are excluded, the average value is
$\sin^2 \theta_W = 0.2227 \pm 0.0004$ \cite{lep01}
in the on-shell scheme. The NuTeV value,
$\sin^2 \theta_W = 0.2277 \pm 0.0013 \, \text{(stat)} 
                         \pm 0.0009 \, \text{(syst)}$  \cite{nutev02},
is significantly larger than the average of the other data.
This difference is called ``NuTeV anomaly". Independent experiments are
in progress to measure the angle by the M$\o$ller scattering \cite{moller}
and parity-violating electron scattering \cite{qweak}. It is also
interesting to find $Q^2$ dependence of $\sin^2 \theta_W$ from various
experiments. From these measurements, we expect that the experimental
situation will become clear. However, since it is an important fundamental
constant, fast clarification of the NuTeV deviation is needed.

There are a number of papers which intended to explain the deviation.
Although there may be new mechanisms beyond the standard model \cite{new},
we should investigate all the possibilities within the present theoretical
framework. Next-to-leading-order effects are not a source of the deviation
\cite{nlo}. There are proposals for conventional explanations \cite{conv}.
First, because the used target is the iron nucleus instead of
the isoscalar nucleon, it is natural to explain it in terms of nuclear
corrections \cite{nutev-nucl,nucl-shadow, nucl-valence, nucl-n-excess}.
However, there is still no clear interpretation at this stage. 
Second, the difference between strange and antistrange quark distributions 
should contribute to the deviation \cite{nutev-prd02,strange,s-cteq,s-nutev}. 
However, it cannot be uniquely determined at this stage. For example,
a CTEQ analysis result \cite{s-cteq} differs from
a NuTeV result \cite{s-nutev}. Third, there are proposals that
the anomaly could be explained by isospin violation \cite{isospin},
which could be related to the nuclear correction in Ref. \cite{nucl-valence}
as shown in Sec. \ref{effect}.
 
In this paper, we investigate a conventional explanation in terms of
a nuclear correction. We study the details of the nuclear effect in
Ref. \cite{nucl-valence}, where it was pointed out that the nuclear
modification difference between the $u_v$ and $d_v$ distributions
contributes to the anomaly. Because such a nuclear modification is unknown,
possible differences are simply estimated by considering
baryon-number and charge conservations \cite{nucl-valence}.
It is the purpose of this paper to clarify this kind of nuclear
effect by a $\chi^2$ analysis with many nuclear data by assigning
explicit parameters for the difference. We also estimate uncertainties
of the difference by the Hessian method, and they are especially
important for discussing an effect on the $\sin^2 \theta_W$ anomaly. 

This paper is organized as follows. In Sec. \ref{pw}, we show
a nuclear modification effect on $\sin^2 \theta_W$ by using
the Paschos-Wolfenstein relation. Next, a $\chi^2$
analysis method is explained in Sec. \ref{analysis} for determining
the modification difference between $u_v$ and $d_v$. Analysis results and
an actual effect on the NuTeV $\sin^2 \theta_W$ are discussed in
Sec. \ref{results}. These studies are summarized in Sec. \ref{summary}.

\section{Corrections to Paschos-Wolfenstein relation}\label{pw}

Neutrino and antineutrino scattering data are used for extracting
the weak mixing angle by the NuTeV collaboration. In discussing
various corrections to the NuTeV measurement, it is theoretically
useful to find the corrections to the Paschos-Wolfenstein (PW)
relation \cite{pw}:
\begin{equation}
R^-  = \frac{  \sigma_{NC}^{\nu N}  - \sigma_{NC}^{\bar\nu N} }
              {   \sigma_{CC}^{\nu N}  - \sigma_{CC}^{\bar\nu N} }
        =  \frac{1}{2} - \sin^2 \theta_W 
\, ,
\label{eqn:pw}
\end{equation}
where $\sigma_{CC}^{\nu N}$ and $\sigma_{NC}^{\nu N}$ are
charged-current (CC) and neutral-current (NC) cross sections,
respectively, in neutrino-nucleon deep inelastic scattering.

The various correction factors to the PW relation are discussed in
Ref. \cite{nucl-valence}. Because they are important for our analysis,
an outline is discussed in this section. 
The iron target is used in the NuTeV experiment, so that nuclear
corrections should be properly taken into account. 
The correction investigated in this paper is the nuclear
modification difference between the $u_v$ and $d_v$ distributions. 
Nuclear modifications of the valence-quark distributions, $u_v$ and $d_v$,
are expressed by $w_{u_v}$ and $w_{d_v}$, which are defined by
\begin{align}
\! \! \! \! 
u_v^A (x,Q^2) & = w_{u_v} (x,Q^2,A,Z) \, \frac{Z u_v (x,Q^2) 
                                         + N d_v (x,Q^2)}{A},
\nonumber \\
\! \! \! \! 
d_v^A (x,Q^2) & = w_{d_v} (x,Q^2,A,Z) \, \frac{Z d_v (x,Q^2) 
                                         + N u_v (x,Q^2)}{A}.
\label{eqn:wpart}
\end{align}
Here, $u_v^A$ and $d_v^A$ are the up- and down-valence quark distributions,
respectively, in a nucleus; $u_v$ and $d_v$ are the distributions
in the proton. The variable $x$ is the Bjorken scaling variable,
$Q^2$ is defined by the momentum transfer $q$ as $Q^2=-q^2$, 
$Z$ is the atomic number of the target nucleus, and $A$ is its mass number.
In Ref. \cite{npdf04}, the weight functions $w_{u_v}$ and $w_{d_v}$ are
defined only at the fixed $Q^2$ point, $Q^2$=1 GeV$^2$; however,
they are defined at any $Q^2$ in the present research. 
As mentioned in Ref. \cite{saga01}, the form of Eq. (\ref{eqn:wpart})
is not appropriate at $x \rightarrow 1$ because the nuclear distributions
do not vanish at $x=1$. However, it is more practical at this stage
to use the form due to the lack of large-$x$ data. In any case, the
large $x$ ($>0.8$) region does not contribute to the anomaly because
of the NuTeV kinematics. 
For investigating the difference between the modifications, we define
a function 
\begin{equation}
\varepsilon_v (x) = \frac{w_{d_v}(x,Q^2,A,Z)-w_{u_v}(x,Q^2,A,Z)}
                             {w_{d_v}(x,Q^2,A,Z)+w_{u_v}(x,Q^2,A,Z)}
\, .
\label{eqn:ev}
\end{equation}
The function $\varepsilon_v$ depends also on $A$, $Z$, and $Q^2$,
but these factors are abbreviated for simplicity.

Writing the cross sections in terms of the nuclear parton distribution
functions (PDFs) in the leading order of $\alpha_s$, we obtain
a modified PW relation with small correction factors.
Expanding the modified relation in terms of the correction factors
and retaining only the leading correction of $\ \varepsilon_v$,
we obtain
\begin{align}
& R_A^-   =  \frac{1}{2} - \sin^2 \theta_W  
\nonumber \\
       &  \ 
- \varepsilon_v (x)  \bigg \{ \bigg ( \frac{1}{2} - \sin^2 \theta_W \bigg ) 
               \frac{1+(1-y)^2}{1-(1-y)^2} - \frac{1}{3} \sin^2 \theta_W 
\bigg \}
\nonumber \\
       &  \ 
+O(\varepsilon_v^2)+O(\varepsilon_n)+O(\varepsilon_s)+O(\varepsilon_c) 
\, .
\label{eqn:apw3}
\end{align}
Here, $O(\varepsilon)$ indicates a correction of the order of $\varepsilon$,
and the variable $y$ is given by the energy transfer $q^0$ and
neutrino energy $E_\nu$ as $y=q^0/E_\nu$. The derivation
of Eq. (\ref{eqn:apw3}) is found in Ref. \cite{nucl-valence}. 
The details are investigated for the first correction factor
$\varepsilon_v (x)$ in this paper.
Other corrections come from neutron excess, strange-antistrange 
asymmetry ($s-\bar s$), and charm-anticharm asymmetry ($c-\bar c$).
They are denoted $O(\varepsilon_n)$, $O(\varepsilon_s)$,
and $O(\varepsilon_c)$ in Eq. (\ref{eqn:apw3}).
The $\varepsilon$ factors are defined by 
$\varepsilon_n=(N-Z)(u_v-d_v)/[A(u_v+d_v)]$,
$\varepsilon_s = s_v^A /[w_v \, (u_v+d_v)]$ and
$\varepsilon_c = c_v^A /[w_v \, (u_v+d_v)]$
with $w_v = (w_{d_v}+w_{u_v})/2$ and $q_v^A \equiv q^A - \bar q^{\, A}$.
The correction factor $\varepsilon_v$ is related to the isospin violation
as shown in Sec. \ref{effect}. The studies of these contributions to
$\sin^2 \theta_W$ should be found elsewhere.
In deriving Eq. (\ref{eqn:apw3}), differential cross sections are used
instead of the total ones for taking the PW ratio. The equation is useful
for finding theoretical possibilities; however, it is practically limited
for estimating a numerical effect because NuTeV kinematical conditions
should be taken into account. This point is discussed in Sec. \ref{results}.
Investigating the modified PW relation, we find that the $\varepsilon_v (x)$
factor contributes to the $\sin^2 \theta_W$ measurement in the neutrino
scattering. We explain the distribution $\varepsilon_v (x)$ and
its effect on the $\sin^2 \theta_W$ determination in the following sections.

\section{$\chi^2$ analysis method for determining 
         $\varepsilon_v (x)$}
\label{analysis}

The nuclear modification difference between the distributions 
$u_v$ and $d_v$ is not known at this stage.
There are few theoretical guidelines for calculating
the difference. Of course, the valance-quark distributions are
constrained by the baryon-number and charge conservations, so that
there should be some restrictions to their nuclear modifications.
From this consideration, two possibilities were proposed for
$\varepsilon_v (x)$ in Ref. \cite{nucl-valence}.
However, the conservation conditions are not enough to impose
$x$-dependent shape of these modifications. It is, therefore,
an appropriate way is to show the modification difference by analyzing
available experimental data. 

We express the difference by a number of parameters, which are then
determined by a $\chi^2$ analysis of nuclear data.
The difference between the valence-quark modifications is denoted 
\begin{equation}
\Delta w_v (x,Q^2,A,Z) = w_{u_v} (x,Q^2,A,Z) -  w_{d_v} (x,Q^2,A,Z) ,
\end{equation}
and their average is $w_v$ as used in the previous section. 
The difference at $Q^2$=1 GeV$^2$ ($\equiv Q_0^2$)
is expressed by four parameters, $a_v'$, $b_v'$, $c_v'$, and $d_v'$:
\begin{align}
\Delta w_v (x,Q_0^2, & A,Z) = \left( 1 - \frac{1}{A^{1/3}} \right)
\nonumber \\
  & \times  \frac{a_v' (A,Z) +b_v' x+c_v' x^2 +d_v' x^3}{(1-x)^{\beta_v}}.
\label{eqn:dwv}
\end{align}
For the functions $w_v$, $w_{\bar q}$, and $w_g$, which indicate
nuclear modifications of average valence-quark, antiquark, and gluon 
distributions, we use the ones obtained by a recent global analysis
in Ref. \cite{npdf04}. The reason of selecting the functional form
of Eq. (\ref{eqn:dwv}) is discussed in Ref. \cite{saga01}.
We briefly explain the essential point. The $A$ dependence of 
the modification is assumed to be proportional to $1-1/A^{1/3}$ 
by considering nuclear volume and surface contributions to
the cross section \cite{sd}. It should be noted that such $A$ dependence
could be too simple to describe the distributions.
The factor $1/(1-x)^{\beta_v}$ is introduced
so as to explain the large-$x$ Fermi-motion part. Looking at typical
data of $F_2^A/F_2^D$ ratios, we find that a cubic function
seems to be appropriate for the $x$ dependence. However, we should 
aware that an appropriate functional form is scarcely known
for $\Delta w_v$.

The parameters $a_v'$, $b_v'$, $c_v'$, and $d_v'$ are determined by
analyzing experimental data for structure-function ratios $F_2^A/F_2^{A'}$
and Drell-Yan cross-section ratios $\sigma_{DY}^{pA}/\sigma_{DY}^{pA'}$.
The details of the experimental data sets are discussed in
Ref. \cite{npdf04}. The $F_2^A/F_2^{A'}$ data are from 
the European Muon Collaboration (EMC) \cite{emc88-90-93}, 
the SLAC-E49, E87, E139, and E140 Collaborations \cite{slac83-83B-88-94},
the Bologna-CERN-Dubna-Munich-Saclay (BCDMS) Collaboration \cite{bcdms85-87}, 
the New Muon Collaboration (NMC) \cite{nmc95-96-96snc},
the Fermilab-E665 Collaboration \cite{e665-92-95},
and HERMES Collaboration \cite{hermes03}.
The Drell-Yan data are taken by the Fermilab-E772 
and E866/NuSea Collaborations \cite{e772-90-e866-99}.
These data are for the nuclei: deuteron, helium-4, lithium, beryllium, carbon,
nitrogen, aluminum, calcium, iron, copper, krypton, silver, tin, xenon,
tungsten, gold, and lead. The total number of the data is 951. 
Because the small-$x$ data are taken in the small $Q^2$ region, our 
leading-twist analysis could be affected especially at small $x$
\cite{npdf04}.

One of these parameters is fixed by baryon-number or charge
conservation, and $a_v'$ is selected for this fixed quantity.
In addition, the parameter $\beta_v$ is taken as the same value
in the analysis of Ref. \cite{npdf04} ($\beta_v=0.1$).
Therefore, there are three free parameters in the analysis.
The total $\chi^2$ is defined by 
\begin{equation}
\chi^2 = \sum_j \frac{(R_{j}^{data}-R_{j}^{theo})^2}
                     {(\sigma_j^{data})^2},
\label{eqn:chi2}
\end{equation}
where $R_j$ indicates the ratios, $F_2^A/F_2^{A'}$ and 
$\sigma_{DY}^{pA}/\sigma_{DY}^{pA'}$.
The theoretical ratios $R_{j}^{theo}$ are calculated by evolving
the nuclear PDFs with the valence-quark
modification difference in Eq. (\ref{eqn:dwv}) by
the Dokshitzer-Gribov-Lipatov-Altarelli-Parisi (DGLAP) equations
to experimental $Q^2$ points. The total $\chi^2$ is minimized by
the subroutine {\tt MINUIT} \cite{minuit}.
Running the subroutine, we obtain optimized values for the parameters
and also a Hessian matrix for error estimation.
The uncertainty of the weight function $\Delta w_v (x,Q^2,\hat{\xi})$ is
calculated with the Hessian matrix:
\begin{align}
[\delta \Delta w_v (x,Q^2)]^2 
    = & \Delta \chi^2  \sum_{i,j}
    \left ( \frac{\partial \Delta w_v (x,Q^2,\xi)}{\partial \xi_i}
     \right )_{\xi=\hat\xi}      
\nonumber \\
&     \times  \, H_{ij}^{-1} \,
    \left ( \frac{\partial \Delta w_v (x,Q^2,\xi)}{\partial \xi_j}
     \right )_{\xi=\hat\xi} \, ,
        \label{eq:dnpdf}
\end{align}
where a parameter is denoted $\xi_i$ and the optimized point is denoted 
$\hat \xi$.
The $\Delta \chi^2$ value is taken so that the confidence level
is the one-$\sigma$-error range for the normal distribution \cite{npdf04}.
Because the parameter number is three, we have $\Delta \chi^2=3.527$.
This uncertainty estimation is especially important in discussing
an effect on the NuTeV anomaly.

\section{Results}
\label{results}

\subsection{Valence-quark modification difference \\
            between $u_v$ and $d_v$}
\label{dwv}
            
We explain $\chi^2$ analysis results. From the analysis, we obtain
$\chi^2_{min}/d.o.f.$=1.57. The parameters are defined at $Q^2$=1 GeV$^2$,
and their optimum values obtained by the fit are shown in
Table \ref{table:parameters}. We note that the constant $a_v '$ is fixed
by one of the baryon-number and charge conservations,
and it depends on nuclear species.  Several nuclei
($D$, $C$, $Ca$, $Fe$, $Ag$, $Au$) are selected for listing its values.
It is almost constant from a small nucleus ($a_v ' (D)=0.00612$)
to a large one ($a_v ' (Au)=0.00615$). The parameters $b_v '$, $c_v '$, 
and $d_v '$ are determined by the analysis. We notice huge errors
which are almost an order of magnitude larger than the optimum
parameter values. This fact indicates that the determination of
$\Delta w_v$ is almost impossible at this stage. However, it is
important to show an uncertainty range of $\Delta w_v$ in order to
compare with the NuTeV anomaly. 
            
\begin{table}[h]
\caption{\label{table:parameters}
         Parameters obtained by the analysis.
         The constant $a_v '$ depends on nuclear species.
         Its value is listed for typical nuclei.}
\vspace{0.2cm}
\begin{tabular}{cc} 
\hline\hline
\ \ \ \ parameter \ \ \ \ & 
\ \ \ \ \ \ \ \ \ \ \ \ value \ \ \ \ \ \ \ \ \ \  \ \  \\
\hline
$b_v '$        &   -0.0905  $\pm$ 0.4044                \\
$c_v '$        & \  0.239   $\pm$ 1.300                 \\
$d_v '$        &   -0.193   $\pm$ 1.126                 \\
\hline
$a_v ' (D)$    &   0.00612                              \\
$a_v ' (C)$    &   0.00612                              \\
$a_v ' (Ca)$   &   0.00612                              \\
$a_v ' (Fe)$   &   0.00613                              \\
$a_v ' (Ag)$   &   0.00614                              \\
$a_v ' (Au)$   &   0.00615                              \\
\hline\hline
\end{tabular}
\end{table}

\begin{figure}[t]
\includegraphics[width=0.46\textwidth]{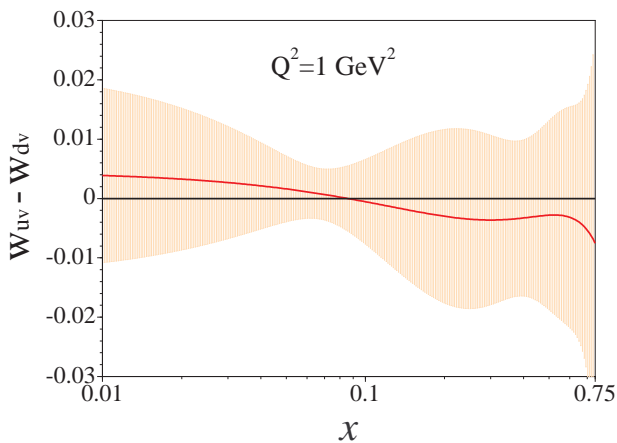}
\vspace{-0.4cm}
\caption{(Color online) 
      Nuclear modification difference $\Delta w_v =w_{u_v}-w_{d_v}$ 
      between $u_v$ and $d_v$ for the iron nucleus. The curve indicates
      the obtained $\Delta w_v$ at $Q^2$=1 GeV$^2$, and the shaded
      area shows the uncertainty range calculated by the Hessian method.}
\label{fig:dwv}
\end{figure}
\vspace{0.0cm}

The difference $w_{u_v}-w_{d_v}$ is plotted as a function of $x$
at $Q^2$=1 GeV$^2$ in Fig. \ref{fig:dwv} by using the optimum parameters.
The solid curve is the distribution obtained by the $\chi^2$ analysis.
The shaded area is the one-$\sigma$ uncertainty range which is calculated
by the Hessian method in Eq. (\ref{eq:dnpdf}). 
Although the distribution is positive at small $x$ and it becomes
negative at large $x$, its functional form is not clear if 
the uncertainties are considered. Because the uncertainty band
is much larger than the distribution itself, we obviously need
future experimental efforts to find an accurate distribution.

\subsection{Effect on $\sin^2 \theta_W$}\label{effect}

\begin{figure}[h]
\includegraphics[width=0.46\textwidth]{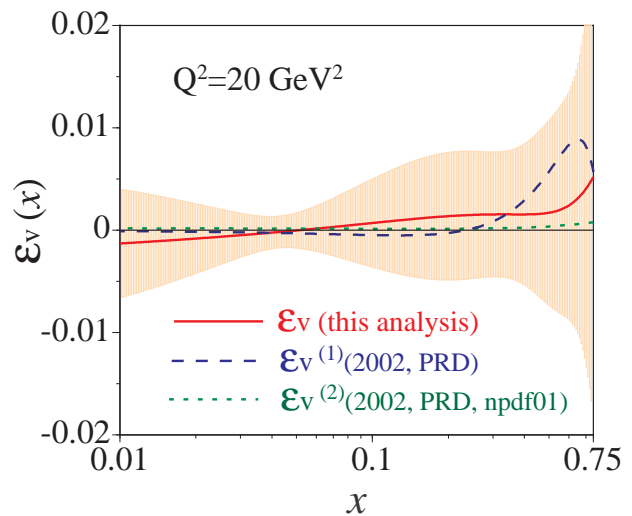}
\vspace{-0.2cm}
\caption{(Color online)
         The function $\varepsilon_v(x)$ at $Q^2$=20 GeV$^2$ 
         for the iron nucleus. 
         The solid curve indicates the current analysis result,
         and the shaded area shows the uncertainty range.
         The dashed and dotted curves indicate the model-1
         and model-2 results, respectively, in Ref. \cite{nucl-valence}.}
\label{fig:ev}
\end{figure}

The average $Q^2$ of the NuTeV neutrino and antineutrino scattering
experiments is about $Q^2$=20 GeV$^2$. The valence-quark distributions
in the iron nucleus are obtained at $Q^2$=1 GeV$^2$ in the previous
section. The nuclear PDFs as well as the nucleonic PDFs are 
evolved to the ones at $Q^2$=20 GeV$^2$ by the DGLAP evolution equations.
Then, the nuclear modifications $w_{u_v}$ and $w_{d_v}$ are 
calculated at $Q^2$=20 GeV$^2$ by using Eq. (\ref{eqn:wpart}),
and the obtained function $\varepsilon_v(x)$ is shown in
Fig. \ref{fig:ev}.

\vspace{0.0cm}

The solid curve is the distribution obtained by the current analysis,
and its uncertainties are shown by the shaded area.
The previous results in Ref. \cite{nucl-valence}
are also shown in the figure. These distributions are proposed
as the ones which satisfy the baryon-number and charge conservations.
The dashed curve indicates the distribution by the model-1, in which 
the conservation integrands are assumed to vanish in the leading
order of the correction factors. The dotted curve indicates
the distribution by the model-2, in which the global $\chi^2$
analysis results for the nuclear PDFs in Ref. \cite{saga01} are used.
Three curves are much different; however, they are certainly
within the error band. It suggests that they should be consistent
within the uncertainties.

Because a Monte-Carlo code, instead of the PW relation, is used
in the experimental analysis for extracting $\sin^2 \theta_W$,
it is not straightforward to compare theoretical results
with the NuTeV deviation. Obviously, NuTeV kinematical
conditions should be taken into account. In particular, few data
exist in the large-$x$ region, where $\varepsilon_v(x)$ is large.
Fortunately, such kinematical factors are provided in
Ref. \cite{nutev-prd02}. In order to use them, our PDFs should be
related to the NuTeV convention \cite{mcfarland}. The up- and
down-valence quark distributions per nucleon are defined by
\begin{align}
x u_v^A = w_{u_v} \, \frac{Z x u_v    +N x d_v}   {A}
        =            \frac{Z u_{vp}^* +N u_{vn}^*}{A} ,
\nonumber \\
x d_v^A = w_{d_v} \, \frac{Z x d_v    +N x u_v}   {A}
        =            \frac{Z d_{vp}^* +N d_{vn}^*}{A} ,
\label{eqn:pdfdef1}
\end{align}
where the middle expressions with $w_{u_v}$ and $w_{d_v}$ are
our notations, and the PDFs with the asterisk (*) are
the NuTeV expressions. Using Eq. (\ref{eqn:pdfdef1}), we find
that the distribution $\varepsilon_v(x)$ is related to
the isospin-violating distributions:
\begin{align}
\delta u_v^* = u_{vp}^*-d_{vn}^* = 
          - \varepsilon_v \, (w_{u_v}+w_{d_v}) \, x \, u_v ,
\nonumber \\
\delta d_v^* = d_{vp}^*-u_{vn}^* = 
          + \varepsilon_v \, (w_{u_v}+w_{d_v}) \, x \, d_v .
\label{eqn:pdfde2}
\end{align}
Therefore, according to the NuTeV convention of the nuclear PDFs,
we have been investigating isospin-violation effects in the nucleon
and their nuclear modifications by the function $\varepsilon_v(x)$.
However, we should aware that such isospin-violation effects could
possibly include other nuclear effects which may not be related to
the isospin violation.
In any case, using the NuTeV functionals in Ref. \cite{nutev-prd02},
we need to calculate the integral
\begin{align}
\Delta (\sin^2 \theta_W) = 
- \int dx & \, \big\{ \, F[\delta u_v^*,x] \, \delta u_v^* (x)
\nonumber \\
             & + F[\delta d_v^*,x] \, \delta d_v^* (x) \, \big\} ,
\label{eqn:del-sinth}
\end{align}
for estimating a correction to the NuTeV measurement.
All the NuTeV kinematical effects are included in the functionals
$F[\delta u_v^*,x]$ and $F[\delta d_v^*,x]$ which are provided
in Fig. 1 of Ref. \cite{nutev-prd02}. It should be noted that
our sign convention of the correction $\Delta (\sin^2 \theta_W)$
is opposite to the NuTeV one.

\begin{figure}[h]
\includegraphics[width=0.46\textwidth]{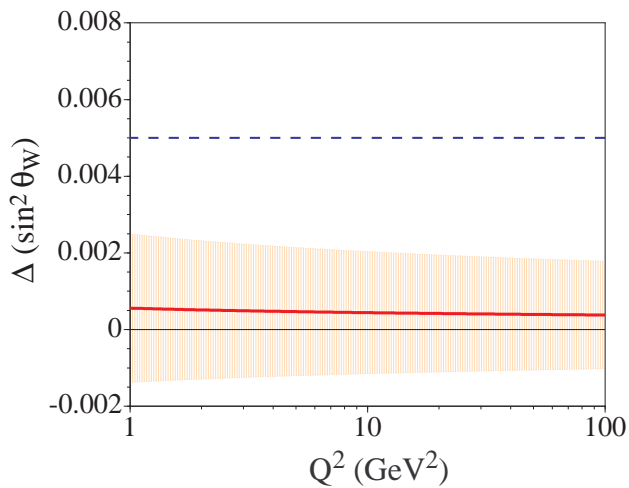}
\caption{(Color online)
         Effects on the NuTeV $\sin^2 \theta_W$ determination 
         as a function of $Q^2$ by taking into account
         the experimental kinematics. 
         The curve indicates the $\chi^2$ analysis result,
         and the shaded area shows the uncertainty range.
         The NuTeV deviation 0.0050 is shown by the dashed line.}
\label{fig:dsinth}
\end{figure}

Using Eqs. (\ref{eqn:pdfde2}) and (\ref{eqn:del-sinth}) together
with the distribution $\varepsilon_v(x)$ obtained in Sec. \ref{dwv},
we calculate the effect on the NuTeV value. 
The results are shown in Fig. \ref{fig:dsinth} as a function of $Q^2$.
The deviation $\Delta (\sin^2 \theta_W)$ is shown by the solid
curve, and the shaded area corresponds to the one-$\sigma$-error
range. For comparison, the NuTeV deviation 0.0050 is shown by the dashed
line in the figure. We notice that the correction is not strongly
dependent on $Q^2$. Even if the uncertainty range is considered,
the contribution is smaller than 0.0050. In fact, calculating
the correction at $Q^2$=20 GeV$^2$, which is approximately
the average $Q^2$ of NuTeV measurements, we obtain
\begin{equation}
\Delta (\sin^2 \theta_W) = 0.0004 \pm 0.0015 .
\label{eqn:effect}
\end{equation}

\vspace{0.0cm}

It indicates that the whole NuTeV anomaly could not be explained
by the nuclear modification difference $w_{u_v}-w_{d_v}$ 
at this stage. However, we should be careful that the uncertainty
estimation depends on the analysis conditions. For example,
a certain $A$ and $x$ dependent functional form is assumed
in the $\chi^2$ analysis and such uncertainties are not included
in Fig. \ref{fig:dsinth} and Eq. (\ref{eqn:effect}). The situation is
the same as the nucleonic PDF case, for example in Ref. \cite{aac03},
where such functional uncertainties are not also estimated.
In addition, the uncertainties due to other parameter errors in Ref. 
\cite{npdf04} are not included. 
There is a possibility that the uncertainty 0.0015 could be underestimated.

Because of these issues, it is, strictly speaking, too early to
exclude the mechanism for explaining the NuTeV anomaly. However,
if the NuTeV anomaly were to be explained solely by the nuclear
modification of the PDFs, the magnitude should be a factor of
ten larger than the global analysis result. Furthermore, the
obtained error is a factor of three smaller than the NuTeV
discrepancy. Therefore, it is unlikely at this stage that
the anomaly is explained by such an effect according to
the analysis of nuclear data. In any case, because the 
difference $w_{u_v}-w_{d_v}$ cannot be determined from
current experimental data, we need future experimental
efforts to find it. Because it is related to nuclear valence-quark
distributions, possibilities are neutrino scattering experiments
at future neutrino facilities such as MINER$\nu$A \cite{minarnua}
and neutrino factories \cite{sk-nufact03}.

\section{Summary}\label{summary}

We have extracted the difference between nuclear modifications of
$u_v$ and $d_v$ by the $\chi^2$ analysis of nuclear $F_2$ and Drell-Yan
data. We found that it is difficult to determine it at this stage
due to the lack of data which are sensitive to the difference.
Such a difference contributes to the NuTeV determination of $\sin^2 \theta_W$
because the iron target was used in the experiment. Taking the NuTeV
kinematics into account, we estimated the effect on the $\sin^2 \theta_W$.
It is not large enough to explain the whole NuTeV deviation. However, 
the nuclear modification difference cannot be accurately determined
at this stage, we need further efforts to find it and its effect on
the NuTeV determination of $\sin^2 \theta_W$.

\begin{acknowledgments}
S.K. was supported by the Grant-in-Aid for Scientific Research from
the Japanese Ministry of Education, Culture, Sports, Science, and Technology.
He thanks K. S. McFarland for suggesting the NuTeV functionals
of Eq. (\ref{eqn:del-sinth}) in order to take the NuTeV kinematical
condition into account in our theoretical analysis.
\end{acknowledgments}



\end{document}